\documentclass[conference]{IEEEtran}
\IEEEoverridecommandlockouts
\usepackage{cite}
\usepackage{amsmath,amssymb,amsfonts,mathrsfs,amsthm}
\usepackage{algorithm}
\usepackage{algpseudocode}
\usepackage{graphicx}
\usepackage{textcomp}
\usepackage{xcolor}
\usepackage{url}
\usepackage{adjustbox}
\usepackage{balance}
\usepackage{subcaption}
\usepackage{makecell}

\begin{document}

\title{Semantic Communication-Empowered Vehicle Count Prediction for Traffic Management}

\author{\IEEEauthorblockN{Sachin Kadam and Dong In Kim}
\IEEEauthorblockA{{Department of Electrical and Computer Engineering} \\
{Sungkyunkwan University (SKKU), Suwon 16419, Republic of Korea}\\
Email: sachinkadam@skku.edu, dikim@skku.ac.kr}
\thanks{This research was supported in part by the MSIT (Ministry of Science and ICT), Korea, under the ICT Creative Consilience program (IITP-2020-0-01821) and the ITRC support program (IITP-2023-RS-2023-00258639) supervised by the IITP (Institute for Information \& Communications Technology Planning \& Evaluation).}
}

\maketitle

\begin{abstract}
Vehicle count prediction is an important aspect of smart city traffic management. Most major roads are monitored by cameras with computing and transmitting capabilities. These cameras provide data to the central traffic controller (CTC), which is in charge of traffic control management. In this paper, we propose a joint CNN-LSTM-based semantic communication (SemCom) model in which the semantic encoder of a camera extracts the relevant semantics from raw images. The encoded semantics are then sent to the CTC by the transmitter in the form of symbols. The semantic decoder of the CTC predicts the vehicle count on each road based on the sequence of received symbols and develops a traffic management strategy accordingly. Using numerical results, we show that the proposed SemCom model reduces overhead by $54.42\%$ when compared to source encoder/decoder methods. Also, we demonstrate through simulations that the proposed model outperforms state-of-the-art models in terms of mean absolute error (MAE) and mean-squared error (MSE).

\end{abstract}

\section{Introduction} \label{Sec:Intro}
\begin{figure}
\centering
\includegraphics[width=0.485\textwidth]{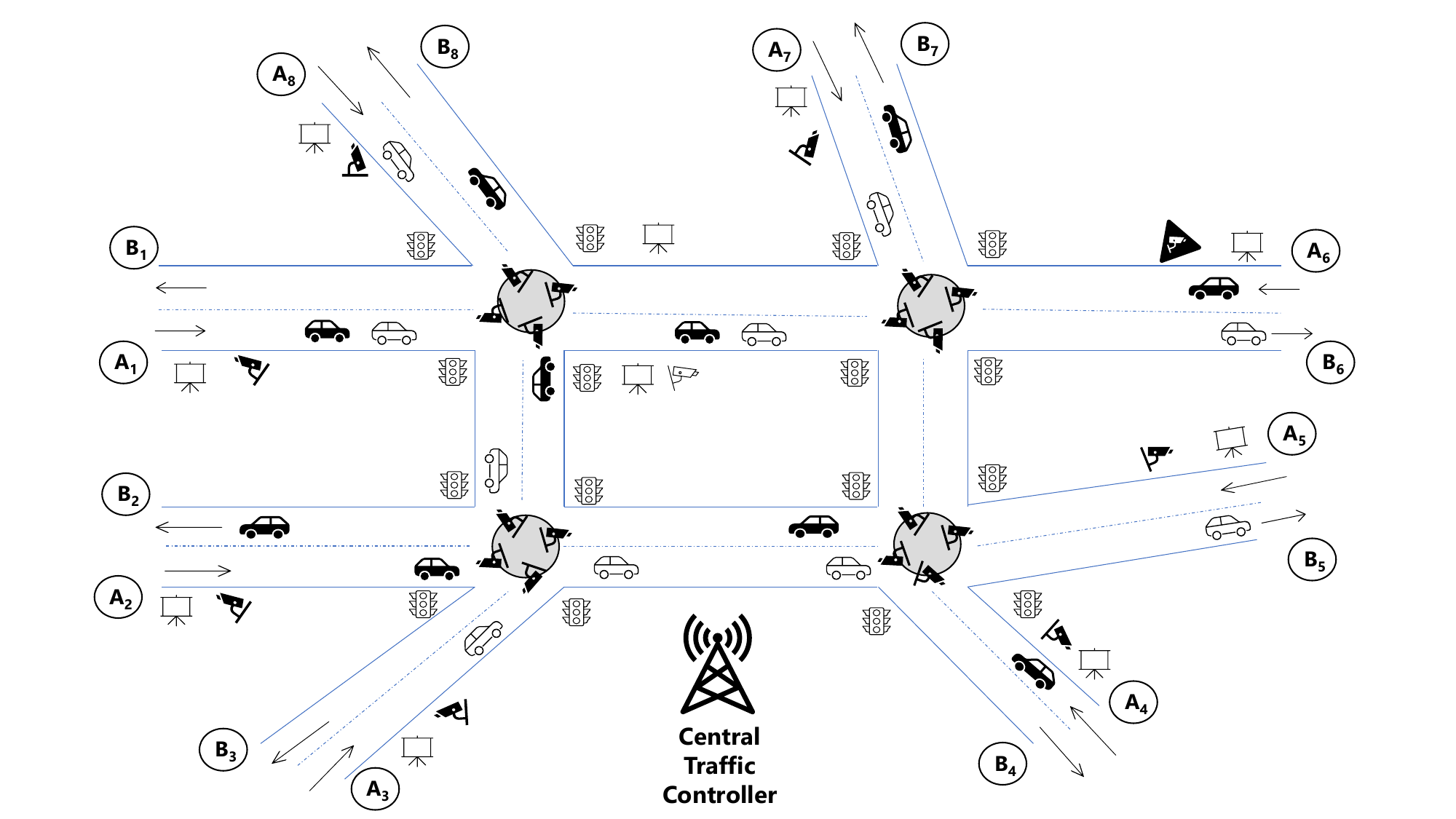}
    \caption{\small The sample traffic model of a smart city with entry ($A_i, i \in \{1, \ldots, 8\}$) and exit ($B_j, j \in \{1, \ldots, 8\}$) points. The main purpose of traffic management is to minimize the average travel time of a traveler to cross a city, i.e., the average time taken between an entry point $A_i, i \in \{1, \ldots, 8\}$ and an exit point $B_j, j \in \{1, \ldots, 8\}$.}
    \label{fig:TrafficModel}
    \vspace{-.4cm}
\end{figure}

The efficient management of vehicular traffic is a key problem in smart city projects. 
The central traffic controller (CTC), which is responsible for traffic control management, requires real-time information on the vehicle density on all major roads.\footnote{Suppose there is heavy traffic on road A, and the traffic controller releases traffic on road B, leading these vehicles to road A, the average waiting time increases. Instead, it can divert some of this traffic to another road C in order to reduce the average waiting time.}
The actions of the CTC to control traffic movement can include traffic holding, releasing, diverting, etc. The CTC requires information regarding the vehicles, including the estimated count, location, size, etc. The CTC devises an optimal strategy for efficient traffic management using the information collected from all camera devices, and the actions required to implement this strategy are communicated to users via traffic lights and giant display screens.\footnote{The CTC communicates with the traffic lights and giant display screens via wired networks.}
A typical traffic model of a smart city with several entry and exit points is shown in Fig.~\ref{fig:TrafficModel}. 
In any traffic management system of a smart city, the purpose is to minimize the average travel time of a traveler to cross a city, i.e., the average time taken between entry and exit points. The average traveling time includes the sum of the time taken while driving and the waiting time at the traffic signals. For this purpose, a fast and accurate prediction of vehicle counts across all major roads is important. 

Several cameras are strategically placed to capture a sequence of images of moving vehicles on the roads. These cameras are part of devices that have computing and transmitting capabilities. The captured raw images are large in size, and transmitting them would add unnecessary overhead to the important data while also increasing the burden on network traffic. Furthermore, the CTC discards irrelevant information from the images received from these devices. This motivates us to propose a semantic communication (SemCom)-based vehicle count prediction model that is fast and accurate. We discuss the proposed model in detail in Section~\ref{Sec:SysModel}. 

Semantic communication is a process that involves transmitting only the information that is relevant to a specific task or job to the intended recipient, resulting in a highly efficient and intelligent system with significantly reduced data traffic~\cite{qin2021semantic}. SemCom systems have attracted a research boom in recent times due to their wide applications in the context of text~\cite{xie2021deep,kadam2023knowledge, kadam2022knowledge}, image~\cite{kang2022personalized,lokumarambage2023wireless}, speech~\cite{han2022semantic,weng2021semantic}, and video~\cite{wang2022wireless, jiang2022wireless}, transmissions. In the context of image transmission, rather than transmitting the entire image as bit sequences, a SemCom-based transmitter extracts only the crucial elements from the source that are necessary for identifying the objects, such as vehicles in the context of traffic management. Extraneous information, such as the sky background, shadows, vegetation, buildings, etc., is eliminated to reduce the amount of data to be transferred without compromising the accuracy of prediction. As a result, there is a significant drop in demand for both power and wireless resources. This results in a more sustainable communication network.

The semantic encoders for image data are typically designed to generate a high-dimensional vector representation of an image that captures its semantic meaning or structure. The most popular semantic encoders for image data include convolutional neural networks (CNNs)~\cite{goodfellow2016deep}.  CNNs are a type of neural network that have proven to be effective for image recognition and classification tasks. They are also shown to perform well even if the sequences of images have low frame rates, heavy occlusion, poor resolution, and so on~\cite{zhang2017fcn}. CNNs are designed to extract features from images at various levels of abstraction automatically, and the resulting feature maps can be used as a semantic representation of the image. Its potential benefits are numerous, including lowered network traffic, limited transmission data overhead, reduced computing complexities at the CTC, and so on. Long short-term memory networks (LSTMs)-based semantic decoders are preferred in the receiver to leverage the temporal correlations between the sequence of images. 

The contributions of this paper are as follows:
\begin{itemize}
    \item We proposed semantic encoder and semantic decoder architectures which make use of CNN's ability to predict at the pixel level and the LSTM's expertise in learning complex temporal dynamics, respectively. 
    \item The proposed joint CNN-LSTM-based semantic encoder-decoder model improves feature representation and allows for an end-to-end trainable mapping from pixels to the prediction of vehicle density, resulting in a novel approach to the problem of vehicle count prediction.
    \item To validate the benefits of our proposed architecture, we use a real-world public dataset to compare vehicle counting performance with state-of-the-art techniques.
    \item We numerically show that the partial residual connection is preferred over full residual connection at the receiver. 
\end{itemize}

The organization of the paper is as follows: A brief literature review on SemCom technologies is provided in Section~\ref{Sec:RelatedWork}. We introduce our proposed joint CNN-LSTM-based system model in Section~\ref{Sec:SysModel}. In Section~\ref{Sec:Prop_Arch}, we present the architectures of the Semantic Encoder and Semantic Decoder models. Next, we provide a few useful simulation results in Section~\ref{Sec:Simulations}. Finally, we conclude the paper in Section~\ref{Sec:Conclusions}.

\section{Related Work} \label{Sec:RelatedWork}
Traffic management in smart cities using Intelligent Transportation Systems (ITS) is a well-studied problem in the literature~\cite{papageorgiou2007its}.  ITS provides robust solutions for real-time traffic network monitoring, prediction, and actuation. 
Recently, due to the emergence of SemCom systems, the implementation of traffic control methods has improved~\cite{raha2023artificial}. A survey paper for a comparison of various traffic models is discussed in~\cite{storani2021analysis}. Semantic feature extraction is a crucial part of any SemCom-based system. For this purpose, based on the context, several deep learning approaches are used for feature extraction. 

The vehicle counting prediction models are designed in several works~\cite{kilic2021accurate,zhao2022vehicle,jin2022dense,hu2022wsnet,hu2022skt,cao2022ghostcount,guo2022dense,sawah2023accurate,xu2022efficient}. A simple and effective single-shot detector model for detecting and counting cars from stationary images is proposed in~\cite{kilic2021accurate}. Another vehicle counting model that takes advantage of cross-resolution spatial consistency and intra-resolution time regularity restrictions is proposed in~\cite{zhao2022vehicle}. But these approaches work effectively when there is no temporal interrelation between the sequence of images. The synergistic attention network (SAN)-based vehicle counting approach is proposed in~\cite{jin2022dense} wherein this method performs dense counting assignments by combining the benefits of transformers and convolutions. A method for estimating local-global traffic density based on weakly supervised learning (WSNet) is proposed in~\cite{hu2022wsnet}. Based on structured knowledge transfer, a lightweight traffic density estimation method (Le-SKT) is proposed in~\cite{hu2022skt}. GhostCount, a lightweight CNN, designed specifically for high-accuracy vehicle counts on edge devices, is proposed in~\cite{cao2022ghostcount}. First, they combined ResNet-18 and Lightweight RefineNet network architectures capable of extracting vehicle features in complex traffic scenes, and then they replaced the regular convolutional layers in ResNet-18 with Ghost modules. For dense traffic detection at highway-railroad crossings, a dense traffic detection net (DTDNet) is developed in~\cite{guo2022dense}. The vehicle counting methods using only gated recurrent unit (GRU) and, also only LSTM are proposed in~\cite{sawah2023accurate}.  Based on a cooperative learning framework, two vehicle counting approaches are proposed in~\cite{xu2022efficient}. Apart from vehicle counting, deep learning models are designed in different contexts like crowd counting~\cite{li2018csrnet,wang2020distribution,yan2021crowd}, object counting~\cite{moreu2022domain,cheng2022rethinking}, crowd-and-vehicle counting~\cite{yu2022frequency}, and so on.

However, most of these vehicle count prediction models have low accuracy, are not robust to vehicle movement, do not capture real-time data, are dependent on accurate data, have high latency in acquiring data, and so on. The most important aspect of any traffic management system is the ability to predict the vehicle count quickly and accurately, which is the focus of this paper.

\section{System Model}\label{Sec:SysModel}
\begin{figure}
\centering
\includegraphics[width=0.485\textwidth]{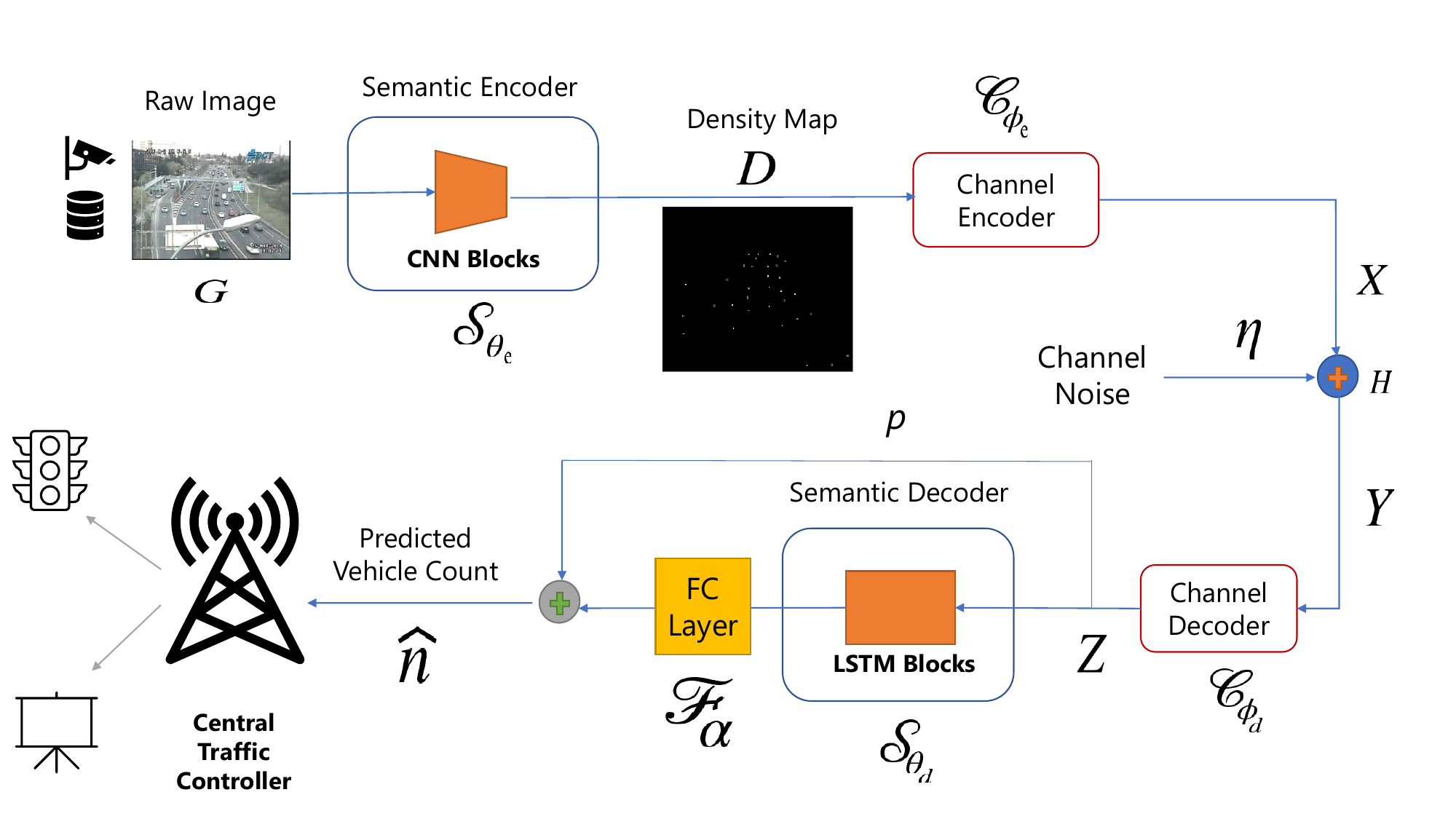}
    \caption{\small The block diagram of our proposed joint CNN-LSTM-based SemCom system model.}
    \vspace{-.4cm}
\label{fig:NetworkModel}
\end{figure} 
\begin{figure*}
\centering
\includegraphics[width=0.925\textwidth]{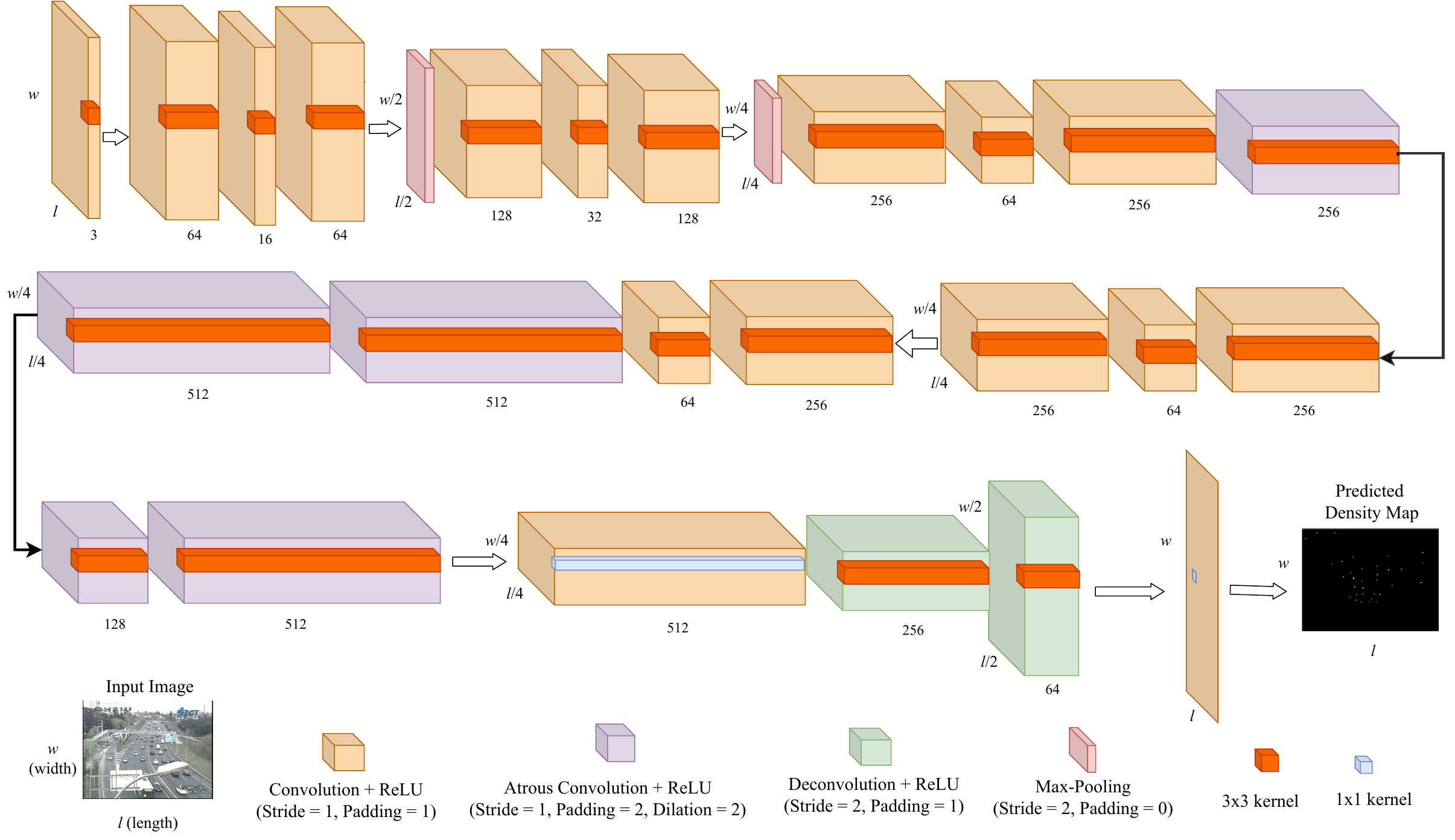}
    \caption{\small The architecture of the semantic encoder, consisting of a CNN that produces the density map whose size is same as that of the input image.}
    \vspace{-.5cm}
\label{fig:CNNBlocks}
\end{figure*} 
The block diagram of our proposed SemCom system model is shown in Fig.~\ref{fig:NetworkModel}. In this system model, camera devices with computing and transmitting capabilities capture raw images. The semantic encoder then extracts the relevant semantics (density maps in this context) from them. Channel encoders then convert them into symbols for transmission. Before reaching the receiver, the symbols are corrupted by channel noise. The channel and semantic decoders predict density maps from the received symbols, which are then used for vehicle counting and traffic control management. 
We exploit the strengths of convolutional neural networks (CNNs) in the semantic encoder for dense visual prediction and of long short-term memory networks (LSTMs) in the semantic decoder for modeling temporal correlation, which are coupled in a partial residual learning framework. At the transmitter, first, the density maps ($D$) are extracted from the sequence of input raw images ($G$) in the semantic encoder using a CNN shown in Fig.~\ref{fig:CNNBlocks}. That is 
\begin{equation}
    D = \mathscr{S}_{\theta_e}(G),
\end{equation}
where $\mathscr{S}_{\theta_e}$ denotes the semantic encoder (see Section~\ref{Sec:SemEncoder}) parameterized by $\theta_e$.
This process keeps only task-relevant information in the data, resulting in a substantial reduction in the overhead\footnote{An example to show this reduction is provided in Section~\ref{Sec:Simulations}.}.    
Next, these density maps are encoded into symbols using the channel encoder, $\mathscr{C}_{\phi_e}$, parameterized by $\phi_e$. After encoding $D$, we get the following set of symbols: $X = \mathscr{C}_{\phi_e} (D)$.

The encoded set of symbols $X$ is transmitted via the AWGN (additive white Gaussian noise) channel. The channel must allow back-propagation for end-to-end training of the semantic encoder and decoder blocks. Simple neural networks, which may create physical channels, for example, are used to represent the AWGN channel.  Let $H$ be the channel gain and $\eta$ be the noise that gets added to $X$ during transmission. Then, the set of received symbols at the receiver is $Y = HX + \eta$. 

After receiving, this set of symbols is decoded using the channel decoder $\mathscr{C}_{\phi_d}$, parameterized by $\phi_d$. After decoding $Y$, we get the following density maps
\begin{equation}
    Z = \mathscr{C}_{\phi_d}(Y).
\end{equation}
Now, to explore the temporal correlations between the sequence of density maps, the LSTM cells are utilized in the semantic decoder (see Section~\ref{Sec:SemDecoder}). Also, instead of directly connecting the residual, as in ResNet~\cite{he2016deep}, a partial residual connection is preferred. 
For this purpose, we introduce a hyper-parameter $p \in [0,1]$, which is multiplied with $Z$ before the addition with the output of semantic decoders $\mathscr{S}_{\theta_d}$, parameterized by $\theta_d$. The output of the semantic decoder is passed through a fully connected (FC) layer, parameterized by $\alpha$, for vehicle count prediction. Hence, the vehicle count prediction is formulated as:
\begin{align}
    \widehat{n} = \mathscr{F}_{\alpha}(\mathscr{S}_{\theta_d}(Z)) + p Z.
\end{align}

\section{Proposed Architectures of Semantic Communication Models}\label{Sec:Prop_Arch}
Now, in subsections~\ref{Sec:SemEncoder} and~\ref{Sec:SemDecoder}, we present the architectures of SemCom models, namely the Semantic Encoder and Semantic Decoder models.
\subsection{Semantic Encoder} \label{Sec:SemEncoder}
The architecture of the designed semantic encoder using a CNN is shown in  Fig.~\ref{fig:CNNBlocks}. 
Existing object counting methods~\cite{onoro2016towards,zhang2016single}, first estimate the object density map of one image and then directly add the density of every pixel in the image. These works motivated our feature extraction method which is designed to generate image density maps for obtaining the vehicle count.
We input the images of size $\ell \times w$, where $\ell$ and $w$ denote the length and width, respectively. The filters are applied such that the output density also has the size $\ell \times w$. The kernels of size $3\times3$ are applied to both convolution and deconvolution layers, as inspired by the VGG-net~\cite{simonyan2014very}. To compensate for the loss of spatial information caused by max pooling, the number of filter channels in the higher layers is increased. In order to reduce the number of parameters, the small filter channels are sandwiched between convolution blocks in a few higher layers. 

Next, we use Atrous convolution, which is equivalent to filter upsampling, by inserting holes between nonzero filter taps. It computes feature maps more densely, then performs simple bilinear interpolation of the feature responses back to the original image size. In comparison to regular convolution, atrous convolution effectively increases the field of view of filters without increasing the number of parameters.  Following that, a convolution layer with $1\times1$ kernels is added to perform feature re-weighting in order to encourage the weighted feature volume to distinguish between foreground and background pixels better. The input of the deconvolution network, which has two deconvolution layers, is the combined and re-weighted feature volume. Lastly, a convolution layer with $1\times1$ kernels acts as a predictor to map features into a vehicle density map.
\subsection{Semantic Decoder} \label{Sec:SemDecoder}
\begin{figure}
\centering
\includegraphics[width=0.48\textwidth]{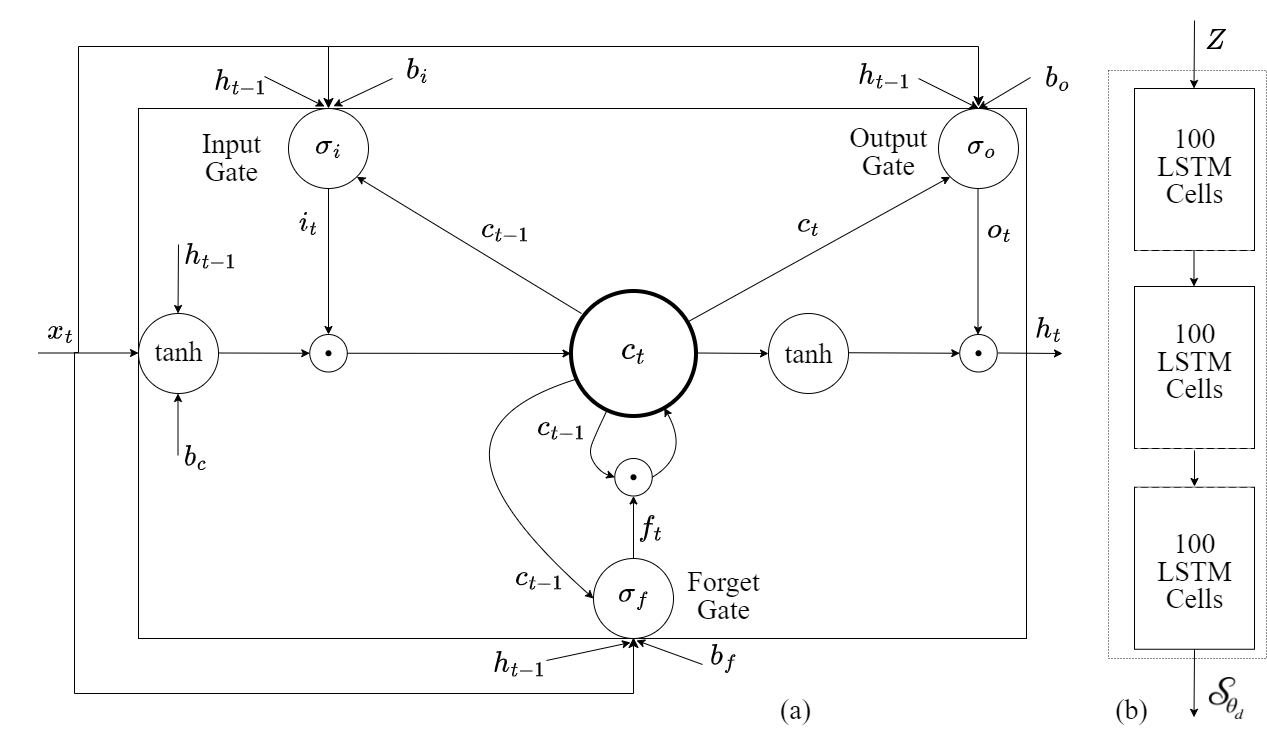}
    \caption{\small (a) Schematic diagram of a single LSTM cell. (b) The architecture of the semantic decoder, consisting of three layers of 100 LSTM cells.}
    \vspace{-.4cm}
\label{fig:LSTMBlocks}
\end{figure} 
In order to incorporate the temporal correlation of vehicle counts from sequential frames, LSTM cells are used to jointly learn vehicle density and vehicle count. The schematic diagram of a single LSTM cell is shown in Fig.~\ref{fig:LSTMBlocks}(a). An LSTM cell consists of three gates: a forget gate $f_t$, an input gate $i_t$, and an output gate $o_t$. These gates allow LSTM to learn and optimize long-term dependencies in succession. Also, LSTM successfully addresses the gradient vanishing/exploding concerns that frequently arise during recurrent neural network training.  It also includes the cell activation vector $c_t$ and the hidden output vector $h_t$. The input gate uses sigmoid function $\sigma_i$ with weight parameters $W_{xi},W_{hi},W_{ci}$, and bias $b_i$; the forget gate uses sigmoid function $\sigma_f$ with weight parameters $W_{xf},W_{hf},W_{cf}$, and bias $b_f$; the output gate uses sigmoid function $\sigma_o$ with weight parameters $W_{xo},W_{ho},W_{co}$, and bias $b_o$.  Also, $\tanh$ functions are used with weight parameters $W_{xc}$, $W_{hc}$, and bias $b_c$. The update equations are obtained as follows (see Fig.~\ref{fig:LSTMBlocks}(a))~\cite{graves2013generating}: 
\begin{subequations}
\begin{align} 
& i_{t}=\sigma_{i}(x_{t}W_{xi}+h_{t-1}W_{hi}+c_{t-1}W_{ci}+b_{i}),\\ 
& f_{t}=\sigma_{f}(x_{t}W_{xf}+h_{t-1}W_{hf}+c_{t-1}W_{cf}+b_{f}),\\ 
& c_{t}=f_{t}\odot c_{t-1}+i_{t}\odot \tanh(x_{t}W_{xc}+h_{t-1}W_{hc}+b_{c}),\\ & o_{t}=\sigma_{o}(x_{t}W_{xo}+h_{t-1}W_{ho}+c_{t}W_{co}+b_{o}),\\ 
& h_{t}=\sigma_{t}\odot \tanh{(c_{t})},
\end{align}
\end{subequations}
where $\odot$ denotes the element-wise product. 

Next, the received density map $Z$ is converted into a one-dimensional vector $x_t$ and fed into the semantic decoder, which consists of three layers of 100 LSTM cells as shown in Fig.~\ref{fig:LSTMBlocks}(b). The total of the learned density map over each frame is used as a base count, and the output hidden vector of the last LSTM layer is input into one fully connected (FC) layer, parameterized by $\alpha$, to learn the difference between the ground truth count and the final estimated count. Numerically, we observed that the partial (p) residual connection simplifies training and improves counting accuracy as compared to the residual connection (see Fig.~\ref{fig:plots_Loss_MAE}(b)).
\subsection{Model Training} \label{Sec:ModelTrain}
The procedure to train the proposed joint CNN-LSTM-based SemCom model is shown in Algorithm~\ref{alg:training}. 
\begin{algorithm}
\caption{Joint CNN-LSTM-based SemCom model training algorithm}\label{alg:training}
\begin{algorithmic}
\State \textbf{Input:} $N, \{G_i,n^0_i, i \!=\! 1, .., N\},p, \lambda, H, K, \overline{T}$, $\eta \! \sim  \!AWGN$ 
\State Initialize $k=1$, $\mathscr{L}_{count} = \overline{T}$
\While{$k \le K$ OR $\mathscr{L}_{count} < \overline{T}$} \Comment{Iterate over $K$ epochs}
\State Initialize $i=1$
\While {$i \le N$} \Comment{Iterate over $N$ batches}
\State $D_i \gets \mathscr{S}_{\theta_e}(G_i)$ \Comment{Semantic Encoder}
\State $\Delta_i \gets \sum_{m=1}^M \left(D_i(m) - D^0_i(m) \right)^2$
\State $X_i \gets \mathscr{C}_{\phi_e}(D_i)$ \Comment{Channel Encoder}
\State $Y_i \gets HX_i + \eta$
\State $Z_i \gets \mathscr{C}_{\phi_d}(Y_i)$ \Comment{Channel Decoder}
\State $\widehat{n}_i \gets \mathscr{F}_{\alpha}(\mathscr{S}_{\theta_d}(Z_i)) + p Z_i$ \Comment{Semantic Decoder}
\State $i \gets i+1$
\EndWhile
\State $\mathscr{L}_{Enc} \gets \frac{1}{N} \sum_{i=1}^N \Delta_i$ \Comment{Sem. Encoder Loss}
\State $\mathscr{L}_{Dec} \gets \frac{1}{N} \sum_{i=1}^N \left(\widehat{n}_i - n^0_i \right)^2$ \Comment{Sem. Decoder Loss}
\State $\mathscr{L}_{count} \gets \mathscr{L}_{Enc} + \lambda \mathscr{L}_{Dec}$ \Comment{Overall Loss}
\State $k \gets k+1$
\EndWhile
\State \textbf{Output:} $\mathscr{S}_{\theta_e}, \mathscr{C}_{\phi_e},
\mathscr{S}_{\theta_d},  \mathscr{C}_{\phi_d}, \mathscr{F}_{\alpha}$.
\end{algorithmic}
\end{algorithm}
For each frame, the semantic encoder predicts the pixel-level density map, while the semantic decoder predicts the vehicle count. These two goals are accomplished in tandem by training the joint CNN-LSTM-based SemCom network end-to-end. The vehicle density is predicted from the feature map using the final $1\times1$ convolution layer of the semantic encoder (see Fig.~\ref{fig:CNNBlocks}). The loss function for density map estimation at the semantic encoder is as follows:
\begin{align}
    \Delta_i &= \sum_{m=1}^M \left(D_i(m) - D^0_i(m) \right)^2, \\
    \mathscr{L}_{Enc} &= \frac{1}{N} \sum_{i=1}^N \Delta_i,
\end{align}
where $D_i(m)$ and $D^0_i(m)$ denote the predicted density map and ground truth density map, respectively, for frame $i$ at pixel $m, \forall m \in \{1, \ldots, M\}$, and $N$ is the batch size. 
Next, the semantic decoder (see Fig.~\ref{fig:NetworkModel}) along with FC layer predicts the vehicle count from the reconstructed density map at the receiver by using the  following expression:  
\begin{equation}
    \widehat{n}_i = \mathscr{F}_{\alpha}(\mathscr{S}_{\theta_d}(Z_i)) + p \sum_{m=1}^M Z_i(m).
\end{equation} 
The squared loss function is used for measuring the vehicle count loss, which is defined as
\begin{equation}
    \mathscr{L}_{Dec} = \frac{1}{N} \sum_{i=1}^N \left(\widehat{n}_i - n^0_i \right)^2,
\end{equation}
where $\widehat{n}_i$ and $n^0_i$ are the predicted vehicle count and ground truth vehicle count, respectively, for frame $i$.
The overall loss $\mathscr{L}_{count}$, used for training the system model is 
\begin{equation}
    \mathscr{L}_{count} = \mathscr{L}_{Enc} + \lambda \mathscr{L}_{Dec},
\end{equation}
where $\lambda$ is a hyper-parameter. 
The batch-based Adam approach~\cite{kingma2014adam}, which is a first-order gradient-based optimization of stochastic objective functions, is used to optimize the loss function.  The model is trained over either a fixed number of $K$ epochs or until the training loss falls below a predetermined threshold value $\overline{T}$. 

\section{Simulation Results}\label{Sec:Simulations}
In this section, the proposed approach is evaluated on the public dataset TRaffic ANd COngestionS (TRANCOS)~\cite{TRANCOSdataset_IbPRIA2015}. It is a benchmark dataset for vehicle density prediction in traffic congestion areas. This dataset comprises 1244 images, with 46796 automobiles labeled in total. All of the pictures were taken using publicly available video surveillance equipment from Spain's Dirección General de Tráfico (\url{https://www.dgt.es/inicio/}). The simulations are performed in a computer with NVIDIA GeForce RTX 3090 GPU and Intel Core i9-10980XE CPU with 256GB RAM. Table~\ref{Tab:Sim_Parameters} shows the simulation parameters used in this paper. 
\begin{table}
\caption{The following simulation parameters are used.}
    \centering
    \resizebox{0.97\columnwidth}{!}{%
    \begin{tabular}{|c|c|c|c|c|c|c|c|}
    \hline
        \makecell{training \\ sample} & \makecell{validation \\ sample} & \makecell{testing \\ sample} & \makecell{learning \\ rate} & \makecell{dropout \\ rate} & epochs & \makecell{batch \\ size} & $\lambda$  \\ \hline
        658 & 165 & 421 & 0.001 & 0.1 & 100 & 8 & 0.001 \\ \hline
    \end{tabular}%
    }
    \label{Tab:Sim_Parameters}
\end{table}
\begin{figure}
\centering
\begin{subfigure}{.24\textwidth}
\centering
\begin{adjustbox}{width = 1\columnwidth}
\includegraphics[width=0.99\textwidth]{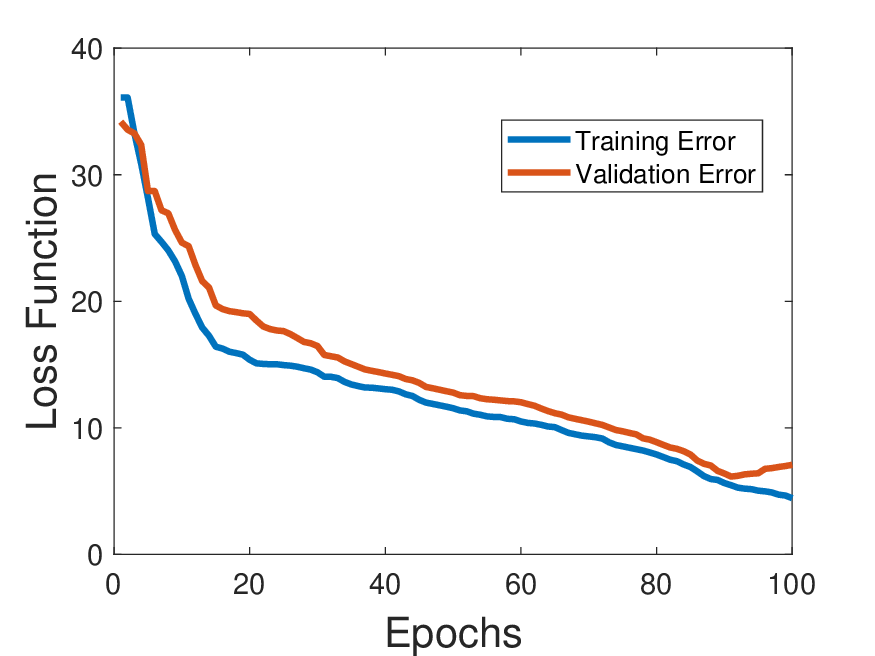}
\end{adjustbox}
\caption{}
\label{fig:Loss_vs_Epoch}
\end{subfigure}%
\begin{subfigure}{.24\textwidth}
\centering
\begin{adjustbox}{width = 1\columnwidth}
\includegraphics[width=0.99\textwidth]{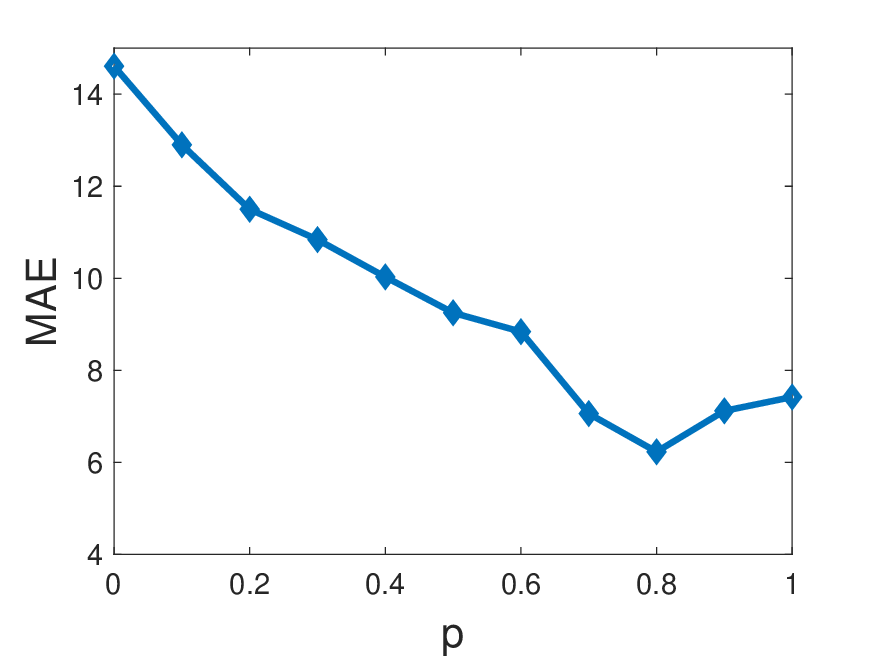}
\end{adjustbox}
\caption{}
\label{fig:MAE_vs_p}
\end{subfigure}
\vspace{-.1cm}
\caption{\small (a) This plot shows the loss function $\mathscr{L}_{count}$ versus the number of epochs for training and validation errors. (b) This plot shows the MAE values versus the hyper-parameter $p$. The minimum MAE value of $6.23$ is obtained when $p=0.8$.}
\vspace{-.1cm}
\label{fig:plots_Loss_MAE}
\end{figure}

Our model is trained by using the procedure described in Algorithm~\ref{alg:training}. For this purpose, the dataset is divided into $4:1$ ratios for training and validation, respectively. The loss function $\mathscr{L}_{count}$ versus the number of epochs for training and validation errors are shown in Fig.~\ref{fig:plots_Loss_MAE}(a). The plot shows that the training and validation errors are close to each other, and after the $95^{th}$ epoch, the validation error begins to deviate from the training error. So, to avoid overfitting and for better results, training is stopped after 100 epochs.\footnote{We repeated the experiment with increased epochs to 150 and checked to see if the validation error had converged and the training error had stabilized. The results confirmed the same conclusion that we drew from Fig.~\ref{fig:plots_Loss_MAE}(a).} 

The accuracy in the predicted value of vehicles is measured in terms of mean absolute error (MAE) and mean-squared error (MSE). They are defined as follows:
\begin{subequations}
\begin{align}
    MAE &= \frac{\sum_{i=1}^{I}|\widehat{n}_i - n^0_i|}{I}, \label{eq:MAE} \\
    MSE &= \frac{\sum_{i=1}^{I}(\widehat{n}_i - n^0_i)^2}{I}, \label{eq:MSE}
\end{align}    
\end{subequations}
where $I$ denotes the number of images in a given frame and $\widehat{n}_i$ (respectively, $n^0_i$) denotes the predicted (respectively, ground truth) vehicle count in the image $i$.
Next, we plot MAE vs. the hyper-parameter $p$, and the results are shown in Fig.~\ref{fig:plots_Loss_MAE}(b). We can deduce from the plot that the minimum MAE value is observed at $p=0.8$, and the corresponding MAE value is 6.23. This also demonstrates that the direct addition of residuals does not always provide the best results. 

In Table~\ref{Tab:MAE_Compare}, the MAE and MSE values computed by the proposed method is compared to three state-of-the-art approaches applied to the TRANCOS dataset. These approaches are based on GRU, LSTM~\cite{sawah2023accurate}, and FCN-rLSTM~\cite{zhang2017fcn}, respectively.  For fair comparison, we replaced our proposed CNN based semantic encoder and LSTM based semantic decoder with those of the described models in~\cite{sawah2023accurate,zhang2017fcn}.  
From the MAE values shown in Table~\ref{Tab:MAE_Compare}, it is deduced that the proposed approach outperforms the approaches based on GRU, LSTM, and FCN-rLSTM, by $90.71\%$, $73.03\%$, and $19.1\%$, respectively. Similarly, in terms of the MSE values, the proposed approach outperforms the approaches based on GRU, LSTM, and FCN-rLSTM, by $103.91\%$, $77.74\%$, and $13.45\%$, respectively.

\begin{table}
\caption{The comparison of the MAE and MSE values between the proposed model and three state-of-the-art models.}
    \centering
    \resizebox{0.95\columnwidth}{!}{%
    \begin{tabular}{|c|c|c|c|c|}
    \hline
        Models & GRU~\cite{sawah2023accurate} & LSTM~\cite{sawah2023accurate} & FCN-rLSTM~\cite{zhang2017fcn} & CNN-LSTM  \\ \hline
        MAE & 11.88 & 10.78 & 7.42 & 6.23 \\ \hline
        MSE & 77.79 & 67.74 & 43.28 & 38.15 \\ \hline
    \end{tabular}%
    }
    \label{Tab:MAE_Compare}
\end{table}

\begin{figure*}
\centering
\begin{adjustbox}{width = 2\columnwidth}
\includegraphics[width=0.99\textwidth]{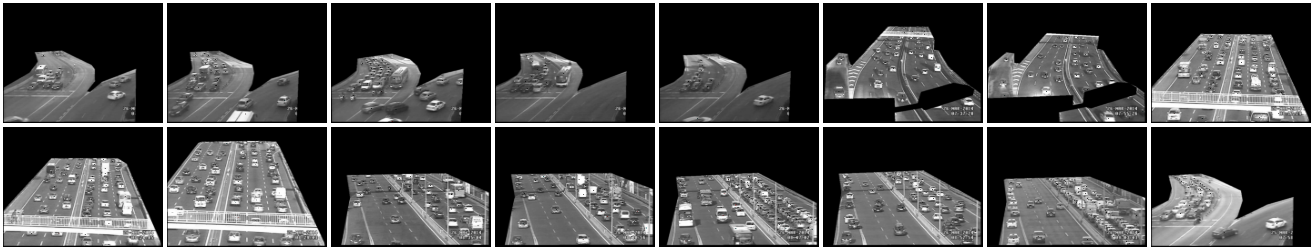}
\end{adjustbox}
\begin{adjustbox}{width = 2\columnwidth}
\includegraphics[width=0.99\textwidth]{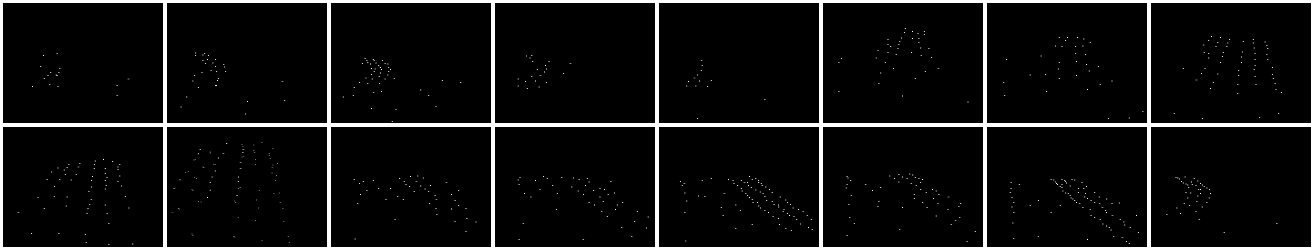}
\end{adjustbox}
\caption{\small The top two rows indicate the raw images taken from the TRANCOS dataset, and the last two rows denote the feature maps extracted from the semantic encoder, respectively.  In this example, the total size of the raw images and transmitted density maps is $195kb$ and $86kb$, respectively. Hence, the overhead reduction between raw data and transmitted data is $55.91\%$.}
\label{fig:Compression}
\vspace{-.4cm}
\end{figure*}

Next, we show how much overhead can be saved by incorporating SemCom technology into our model. As an example, as shown in Fig.~\ref{fig:Compression}, the sizes of raw images from the TRANCOS dataset are compared to the sizes of transmitted density maps from the transmitter. By using the test dataset, which consists of 421 raw images with a total size of $5.31MB$, we compare the overhead reduction with and without applying the SemCom technology. When the semantic encoder/decoder are used, the raw images are compressed to $2.42MB$ (the total size of all 421 encoded images in the test dataset). Hence, a total overhead reduction of $54.42\%$ is achieved by incorporating the SemCom technology for the test dataset.

\section{Conclusions and Future Work}\label{Sec:Conclusions}
In this paper, we propose a joint CNN-LSTM-based SemCom model in which a camera's semantic encoder extracts density maps from raw images. The encoded density maps are then sent as symbols to the CTC by the transmitter. The CTC's semantic decoder predicts the vehicle count on each road based on the sequence of received symbols. We showed that the proposed SemCom model, applied to the TRANCOS dataset, reduces overhead by $54.42\%$ when compared to source encoder/decoder methods using simulations. In addition, simulations confirmed that the proposed model outperforms state-of-the-art models in terms of MAE and MSE. However, the practical deployment challenges of the proposed model need to be addressed. These include hardware requirements for the cameras, the scalability of the system in dense urban environments, and integration with existing traffic management infrastructure. The full operation of traffic management, including deployment challenges, by the CTC can be explored as a future research direction. One approach to the same is presented in~\cite{kadam2023semantic}.

\bibliographystyle{ieeetr}
\balance
\bibliography{references.bib}

\begin{thebibliography}{10}

\bibitem{qin2021semantic}
Z.~Qin, X.~Tao, J.~Lu, W.~Tong, and G.~Y. Li, ``{Semantic Communications: Principles and Challenges},'' {\em arXiv preprint arXiv:2201.01389}, 2021.

\bibitem{xie2021deep}
H.~Xie, Z.~Qin, G.~Y. Li, and B.-H. Juang, ``{Deep Learning Enabled Semantic Communication Systems},'' {\em IEEE Transactions on Signal Processing}, vol.~69, pp.~2663--2675, 2021.

\bibitem{kadam2023knowledge}
S.~Kadam and D.~I. Kim, ``{Knowledge-Aware Semantic Communication System Design},'' in {\em {IEEE} International Conference on Communications, {ICC} 2023, Rome, Italy}, pp.~6102--6107, {IEEE}, 2023.

\bibitem{kadam2022knowledge}
S.~Kadam and D.~I. Kim, ``{Knowledge-Aware Semantic Communication System Design and Data Allocation},'' {\em IEEE Transactions on Vehicular Technology}, Early access, 2023.

\bibitem{kang2022personalized}
J.~Kang, H.~Du, Z.~Li, Z.~Xiong, S.~Ma, D.~Niyato, and Y.~Li, ``{Personalized Saliency in Task-Oriented Semantic Communications: Image Transmission and Performance Analysis},'' {\em IEEE Journal on Selected Areas in Communications}, vol.~41, no.~1, pp.~186--201, 2022.

\bibitem{lokumarambage2023wireless}
M.~U. Lokumarambage, V.~Gowrisetty, H.~Rezaei, T.~Sivalingam, N.~Rajatheva, and A.~Fernando, ``{Wireless End-to-End Image Transmission System Using Semantic Communications},'' {\em IEEE Access}, 2023.

\bibitem{han2022semantic}
T.~Han, Q.~Yang, Z.~Shi, S.~He, and Z.~Zhang, ``{Semantic-Preserved Communication System for Highly Efficient Speech Transmission},'' {\em IEEE Journal on Selected Areas in Communications}, vol.~41, no.~1, pp.~245--259, 2022.

\bibitem{weng2021semantic}
Z.~Weng and Z.~Qin, ``{Semantic Communication Systems for Speech Transmission},'' {\em IEEE Journal on Selected Areas in Communications}, vol.~39, no.~8, pp.~2434--2444, 2021.

\bibitem{wang2022wireless}
S.~Wang, J.~Dai, Z.~Liang, K.~Niu, Z.~Si, C.~Dong, X.~Qin, and P.~Zhang, ``Wireless deep video semantic transmission,'' {\em IEEE Journal on Selected Areas in Communications}, vol.~41, no.~1, pp.~214--229, 2022.

\bibitem{jiang2022wireless}
P.~Jiang, C.-K. Wen, S.~Jin, and G.~Y. Li, ``{Wireless Semantic Communications for Video Conferencing},'' {\em IEEE Journal on Selected Areas in Communications}, vol.~41, no.~1, pp.~230--244, 2022.

\bibitem{goodfellow2016deep}
I.~Goodfellow, Y.~Bengio, and A.~Courville, {\em {Deep Learning}}.
\newblock MIT press, 2016.

\bibitem{zhang2017fcn}
S.~Zhang, G.~Wu, J.~P. Costeira, and J.~M. Moura, ``{FCN-rLSTM: Deep Spatio-Temporal Neural Networks for Vehicle Counting in City Cameras},'' in {\em Proceedings of the IEEE international conference on computer vision}, pp.~3667--3676, 2017.

\bibitem{papageorgiou2007its}
M.~Papageorgiou, M.~Ben-Akiva, J.~Bottom, P.~H. Bovy, S.~P. Hoogendoorn, N.~B. Hounsell, A.~Kotsialos, and M.~McDonald, ``{ITS and Traffic Management},'' {\em Handbooks in operations research and management science}, vol.~14, pp.~715--774, 2007.

\bibitem{raha2023artificial}
A.~D. Raha, M.~S. Munir, A.~Adhikary, Y.~Qiao, S.-B. Park, and C.~S. Hong, ``{An Artificial Intelligent-Driven Semantic Communication Framework for Connected Autonomous Vehicular Network},'' in {\em 2023 International Conference on Information Networking (ICOIN)}, pp.~352--357, IEEE, 2023.

\bibitem{storani2021analysis}
F.~Storani, R.~Di~Pace, F.~Bruno, and C.~Fiori, ``Analysis and comparison of traffic flow models: a new hybrid traffic flow model vs benchmark models,'' {\em European transport research review}, vol.~13, no.~1, pp.~1--16, 2021.

\bibitem{kilic2021accurate}
E.~Kilic and S.~Ozturk, ``An accurate car counting in aerial images based on convolutional neural networks,'' {\em Journal of Ambient Intelligence and Humanized Computing}, pp.~1--10, 2021.

\bibitem{zhao2022vehicle}
Q.~Zhao, J.~Xiao, Z.~Wang, X.~Ma, M.~Wang, and S.~Satoh, ``{Vehicle Counting in Very Low-Resolution Aerial Images via Cross-Resolution Spatial Consistency and Intraresolution Time Continuity},'' {\em IEEE Transactions on Geoscience and Remote Sensing}, vol.~60, pp.~1--13, 2022.

\bibitem{jin2022dense}
Y.~Jin, J.~Wu, W.~Wang, Y.~Wang, X.~Yang, and J.~Zheng, ``{Dense Vehicle Counting Estimation via a Synergism Attention Network},'' {\em Electronics}, vol.~11, no.~22, p.~3792, 2022.

\bibitem{hu2022wsnet}
Y.-X. Hu, R.-S. Jia, Y.-B. Liu, Y.-C. Li, and H.-M. Sun, ``{WSNet: A local--global consistent traffic density estimation method based on weakly supervised learning},'' {\em Knowledge-Based Systems}, vol.~255, p.~109727, 2022.

\bibitem{hu2022skt}
Y.-X. Hu, Q.~Sun, R.-S. Jia, Y.-C. Li, Y.-B. Liu, and H.-M. Sun, ``{Le-SKT: Lightweight traffic density estimation method based on structured knowledge transfer},'' {\em Information Sciences}, vol.~607, pp.~947--960, 2022.

\bibitem{cao2022ghostcount}
Q.~Cao, Z.~Shan, K.~Long, and Z.~Wang, ``{GhostCount: A lightweight convolution network based on high-altitude video for vehicle instantaneous counting in dense traffic scenes},'' {\em IET Intelligent Transport Systems}, 2022.

\bibitem{guo2022dense}
F.~Guo, Z.~Jiang, Y.~Wang, C.~Chen, and Y.~Qian, ``{Dense Traffic Detection at Highway-Railroad Grade Crossings},'' {\em IEEE Transactions on Intelligent Transportation Systems}, vol.~23, no.~9, pp.~15498--15511, 2022.

\bibitem{sawah2023accurate}
M.~S. Sawah, S.~A. Taie, M.~H. Ibrahim, and S.~A. Hussein, ``An accurate traffic flow prediction using long-short term memory and gated recurrent unit networks,'' {\em Bulletin of Electrical Engineering and Informatics}, vol.~12, no.~3, pp.~1806--1816, 2023.

\bibitem{xu2022efficient}
H.~Xu, Z.~Cai, R.~Li, and W.~Li, ``{Efficient CityCam-to-edge cooperative learning for vehicle counting in ITS},'' {\em IEEE Transactions on Intelligent Transportation Systems}, vol.~23, no.~9, pp.~16600--16611, 2022.

\bibitem{li2018csrnet}
Y.~Li, X.~Zhang, and D.~Chen, ``{CSRNet: Dilated Convolutional Neural Networks for Understanding the Highly Congested Scenes},'' in {\em {Proceedings of the IEEE CVPR}}, pp.~1091--1100, 2018.

\bibitem{wang2020distribution}
B.~Wang, H.~Liu, D.~Samaras, and M.~H. Nguyen, ``{Distribution Matching for Crowd Counting},'' {\em Advances in neural information processing systems}, vol.~33, pp.~1595--1607, 2020.

\bibitem{yan2021crowd}
Z.~Yan, R.~Zhang, H.~Zhang, Q.~Zhang, and W.~Zuo, ``{Crowd Counting Via Perspective-Guided Fractional-Dilation Convolution},'' {\em IEEE Transactions on Multimedia}, vol.~24, pp.~2633--2647, 2021.

\bibitem{moreu2022domain}
E.~Moreu, K.~McGuinness, D.~Ortego, and N.~E. O'Connor, ``{Domain Randomization for Object Counting},'' {\em arXiv preprint arXiv:2202.08670}, 2022.

\bibitem{cheng2022rethinking}
Z.-Q. Cheng, Q.~Dai, H.~Li, J.~Song, X.~Wu, and A.~G. Hauptmann, ``{Rethinking Spatial Invariance of Convolutional Networks for Object Counting},'' in {\em Proceedings of the IEEE/CVF Conference on Computer Vision and Pattern Recognition}, pp.~19638--19648, 2022.

\bibitem{yu2022frequency}
X.~Yu, Y.~Liang, X.~Lin, J.~Wan, T.~Wang, and H.-N. Dai, ``{Frequency Feature Pyramid Network With Global-Local Consistency Loss for Crowd-and-Vehicle Counting in Congested Scenes},'' {\em IEEE Transactions on Intelligent Trans. Sys.}, vol.~23, no.~7, pp.~9654--9664, 2022.

\bibitem{he2016deep}
K.~He, X.~Zhang, S.~Ren, and J.~Sun, ``Deep residual learning for image recognition,'' in {\em Proceedings of the IEEE conference on computer vision and pattern recognition}, pp.~770--778, 2016.

\bibitem{onoro2016towards}
D.~Onoro-Rubio and R.~J. L{\'o}pez-Sastre, ``{Towards Perspective-free Object Counting with Deep Learning},'' in {\em Computer Vision--ECCV 2016: 14th European Conference, Amsterdam, The Netherlands, October 11--14, 2016, Proceedings, Part VII 14}, pp.~615--629, Springer, 2016.

\bibitem{zhang2016single}
Y.~Zhang, D.~Zhou, S.~Chen, S.~Gao, and Y.~Ma, ``{Single-image Crowd Counting via Multi-column Convolutional Neural Network},'' in {\em Proceedings of the IEEE conference on computer vision and pattern recognition}, pp.~589--597, 2016.

\bibitem{simonyan2014very}
K.~Simonyan and A.~Zisserman, ``Very deep convolutional networks for large-scale image recognition,'' {\em arXiv preprint arXiv:1409.1556}, 2014.

\bibitem{graves2013generating}
A.~Graves, ``Generating sequences with recurrent neural networks,'' {\em arXiv preprint arXiv:1308.0850}, 2013.

\bibitem{kingma2014adam}
D.~P. Kingma and J.~Ba, ``Adam: A method for stochastic optimization,'' {\em arXiv preprint arXiv:1412.6980}, 2014.

\bibitem{TRANCOSdataset_IbPRIA2015}
R.~Guerrero-G{\'o}mez-Olmedo, B.~Torre-Jim{\'e}nez, R.~L{\'o}pez-Sastre, S.~Maldonado-Basc{\'o}n, and D.~Onoro-Rubio, ``{Extremely Overlapping Vehicle Counting},'' in {\em Pattern Recognition and Image Analysis: 7th Iberian Conference, IbPRIA 2015, Santiago de Compostela, Spain, June 17-19, 2015, Proceedings 7}, pp.~423--431, Springer, 2015.

\bibitem{kadam2023semantic}
S.~Kadam and D.~I. Kim, ``{Semantic Communication-Empowered Traffic Management using Vehicle Count Prediction},'' {\em arXiv preprint arXiv:2307.12254v1}, 2023.

\end{thebibliography}
\end{document}